\documentclass[aip,jcp,amsmath,amssymb,reprint]{revtex4-1}

\usepackage[utf8]{inputenc}
\usepackage[T1]{fontenc}
\usepackage[english]{babel}

\usepackage{geometry}
\usepackage{float}

\usepackage{amsmath}
\usepackage{amssymb}
\usepackage{amsfonts}
\usepackage{mathtools,dsfont,braket}
\usepackage{tensor}
\usepackage{nicefrac}
\usepackage[version=4]{mhchem}
\usepackage{siunitx}

\usepackage{booktabs}
\usepackage{multirow}
\usepackage{graphicx}
\usepackage{dcolumn}
\usepackage{bm}
\usepackage{microtype}
\usepackage{listings,xcolor}
\usepackage{enumerate}
\usepackage[pdftex,colorlinks]{hyperref} 
\hypersetup{%
  linkcolor=blue,%
  citecolor=blue,%
  urlcolor=blue,%
  pdftitle={Limitations of coupled cluster approximations for highly
    accurate investigations of Rb\textsubscript{2}\textsuperscript{$+$}},%
  pdfauthor={Jan Schnabel, Lan Cheng, Andreas K\"ohn}%
}

\usepackage{lipsum}

\newcommand\diff{\mathrm{d}}

\usepackage{mathptmx}
\usepackage{etoolbox}

\makeatletter
\def\@email#1#2{%
 \endgroup
 \patchcmd{\titleblock@produce}
  {\frontmatter@RRAPformat}
  {\frontmatter@RRAPformat{\produce@RRAP{*#1\href{mailto:#2}{#2}}}\frontmatter@RRAPformat}
  {}{}
}%
\makeatother
\begin{document}

\title{Limitations of coupled cluster approximations for highly accurate
  investigations of Rb\textsubscript{2}\textsuperscript{$+$}}

\author{Jan Schnabel}
\affiliation{Institute for Theoretical Chemistry and Center for Integrated
  Quantum Science and Technology, University of Stuttgart, 70569 Stuttgart, Germany}

\author{Lan Cheng}
\affiliation{Department of Chemistry, The Johns Hopkins University, Baltimore,
  Maryland 21218, United States}

\author{Andreas Köhn}
\affiliation{Institute for Theoretical Chemistry and Center for Integrated
  Quantum Science and Technology, University of Stuttgart, 70569 Stuttgart, Germany}
\email{koehn@theochem.uni-stuttgart.de}
\email{lcheng24@jhu.edu}
\email{schnabel@theochem.uni-stuttgart.de}

\date{\today}

\begin{abstract}
  We reveal limitations of several standard coupled-cluster (CC) methods with
  perturbation-theory based noniterative or approximate iterative treatments
  of triple excitations when applied to the determination of highly accurate
  potential energy curves (PECs) of ionic dimers, such as the
  $X{\,}^2\Sigma_g^+$ electronic ground state of \ce{Rb2+}. Such computations
  are of current interest for the understanding of ion-atom interactions in
  the ultracold regime.  We demonstrate that these coupled-cluster methods
  lead to an unphysical long-range barrier for the \ce{Rb2+} system. The
  barrier is small but clearly spoils the entire long-range behavior of the
  PEC. The effect is also found for other \ce{X2+} systems, like \ce{X} $=$
  \ce{Li}, \ce{Na}, and \ce{K}. Calculations using a new flexible framework
  for obtaining leading perturbative triples corrections derived using an
  analytic CC singles and doubles (CCSD) energy derivative formulation
  demonstrate that the origin of this problem lies in the use of $\hat{T}_3$
  amplitudes obtained from approximate CC singles, doubles and triples (CCSDT)
  amplitude equations.  It is shown that the unphysical barrier is related to
  a symmetry instability of the underlying Hartree-Fock mean-field
  solution. Physically meaningful perturbative corrections in the long-range
  tail of the PEC may instead be obtained using symmetry-broken reference
  determinants.

\end{abstract}

\maketitle 

\section{Introduction}
The understanding and the gain of control of interacting neutral atoms in the
ultracold quantum regime has grown substantially over the past decades. The
achievements, to mention only a few of them, reach from Bose-Einstein
condensation (BEC)~\cite{RevModPhys.74.1131}, over Rydberg
systems~\cite{PhysRevX.6.031020,Liebisch_2016,PhysRevA.93.032512} to creating
and controlling ultracold
molecules~\cite{DeissPolarizabilityRb22015,DrewsInealColl2017,Drews2017}. 
While for neutral atoms reaching the ultracold quantum scattering regime (i.e. s-wave collision regime) is nowadays well established, 
it is still a non-trivial challenge to reach the quantum scattering regime for 
hybrid ion-atom systems due to more stringent temperature
requirements~\cite{RevModPhys.91.035001}. 
This is extremely desirable as 
hybrid ion-atom systems are expected to pave the way for novel experiments, phenomena and applications --
among others the precision measurements of ion-atom collision parameters and
associated molecular
potentials~\cite{PhysRevA.62.012709,PhysRevA.79.010702,PhysRevLett.120.153401}.

Novel experimental approaches have been proposed recently in
Refs.~\citenum{Kleinbach2018,Engel2018}. Here the ion-atom
interaction for a core of a giant Rydberg atom immersed in a BEC of
${}^{87}\mathrm{Rb}$ leading to a temperature environment below a
microkelvin has been studied. In principle, the experimental accuracy achievable with this
approach is in the $\mathrm{MHz}$ -- ($=\mathcal{O}(10^{-5}\,\mathrm{cm}^{-1})$
--) domain with a characteristic range of the ion-atom interaction (for
\ce{Rb}) of $R^* = \sqrt{\mu C_4} \approx
5000\,\mathrm{a}_0$~\cite{RevModPhys.91.035001}. 

These pilot experiments, see e.g. Ref.~\citenum{Kleinbach2018} and references therein, aim at entering the s-wave scattering regime
and eventually studying the ro-vibrational structure (of, e.g., the threshold bound
states) and charge-transfer processes of \ce{Rb2+}. Therefore, highly accurate
potential energy curves (PECs) are needed as a starting
point for subsequent studies of corresponding properties related to design
and performance of these experiments. The PECs
have to be accurate not only in the long-range region (up to
$R_\infty=R^*\approx 5000\,\mathrm{a}_0$ to investigate scattering effects
sufficiently), but also in the short-range region (to provide an accurate insight
into the ro-vibrational structure). 
The long-range part of the interaction potential between an
S-state ion and an S-state atom in the electronic ground state is given by~\cite{RevModPhys.91.035001}
\begin{align}
  \label{eq:LR}
  V_{\mathrm{ion-atom}}^{\mathrm{LR}} &\approx -\frac{C_4^{\mathrm{ind}}}{R^4}
  - \frac{C_6^{\mathrm{ind}}}{R^6} - \frac{C_6^{\mathrm{disp}}}{R^6} + \ldots\,.
\end{align}
The $C_4^{\mathrm{ind}}$ and $C_6^{\mathrm{ind}}$ terms describe
the interaction between the charge of the ion and the induced electric dipole
(quadrupole) moment of the atom, while the $C_6^{\mathrm{disp}}$ dispersion
term represents the interaction between instantaneous dipole-induced dipole
moments of the ion and atom arising due to quantum
fluctuations~\cite{RevModPhys.91.035001}. 
Patil and Tang approximately evaluated multipolar polarizabilities and
dispersion coefficients of homonuclear and heteronuclear interactions of both
alkali and alkaline earth atoms and ions, respectively in
Ref.~\citenum{doi:10.1063/1.473089}. This can be further used for models studying
reactive collisions, cf. e.g. Refs.~\citenum{PhysRevLett.110.213202,PhysRevA.90.042705}. 
However, 
using approximate values for the
dispersion coefficient might turn out to be insufficient
for predictions to guide novel experiments.
It is thus of significant interest to obtain accurate PECs using
\emph{ab-initio} calculations.

Theoretical investigations of \ce{X2+}-- systems (with
\ce{X} $=$ \ce{Li}, \ce{Na}, \ce{K}, \ce{Rb})
~\cite{MAGNIER199957,doi:10.1002/qua.24475,MAGNIER2003217,JRAIJ2003129}
have been reported earlier.
As perhaps the only example aiming at high accuracy, 
Tomza {\it{et al.}} reported 
a scheme for obtaining a PEC of 
\ce{Li2+} from relativistic coupled cluster (CC) calculations.%
\cite{PhysRevLett.120.153401} 
Our efforts originally aimed at a first high accuracy calculation 
of PECs for homonuclear molecular ions containing heavier alkali metal species
using an additivity scheme as laid out in Section~\ref{sec:CompAspects}.
Our calculations revealed 
some non-trivial 
subtleties in 
CC methods with
noniterative and approximate iterative treatment of triple excitations including
the coupled-cluster singles and doubles augmented with a noniterative triples correction [CCSD(T)] method, 
-- the `gold standard' of quantum
chemistry. The corresponding theoretical basics are outlined in Sec.~\ref{sec:theory}. As shown in Sec.~\ref{sec:results}, this problem leads to an unphysical barrier in the
long-range region of the PEC.
The present paper thus is focused on understanding and solving this problem.
In Sec.~\ref{sec:proof}  
the problem is analyzed and attributed to a dominant
contribution of the Fockian in the corresponding equations
of these approximate treatments of triple excitations. We show that physically meaningful perturbative corrections in the long-range can be obtained using symmetry-broken reference determinants. 
Moreover, we present an alternative approach with approximate treatment 
of triple excitations to obtain high accuracy and simultaneously avoiding the
long-range problem to extract valuable properties such as dispersion coefficients.
Finally, Section~\ref{sec:ConclusionOutlook} gives a summary 
and an outlook.


\section{\label{sec:CompAspects}Computational aspects}


High-accuracy quantum-chemical calculations of atomic and molecular energies 
often rely on additivity schemes. \cite{Feller93,Helgaker97,Martin99,HEAT2004,Schuurman04,Feller08} 
The present study was originally designed to follow 
the HEAT 
(\textbf{H}igh accuracy \textbf{E}xtrapolated \textbf{A}b initio
\textbf{T}hermochemistry)~\cite{HEAT2004,HEAT2006} protocol 
to obtain as accurate energies as the present computational
resources allow
in a systematic way. Here we assume that the
total electronic energy $E$ of a 
given molecular system can be calculated using the following additivity scheme
\begin{align} 
  \label{eq:HEATorig}
  E &= E_{\mathrm{HF}}^\infty + \Delta E_{\mathrm{CCSD(T)}}^\infty + \Delta
  E_{\mathrm{HLC}} + \Delta E_{\mathrm{higher-rel}}, 
\end{align}
in which $E_{\mathrm{HF}}^\infty$ and $\Delta E_{\mathrm{CCSD(T)}}^\infty$ 
are the estimated complete basis-set (CBS) limit values for the Hartree-Fock (HF) energy 
and the CCSD(T) correlation energy, 
and $E_{\mathrm{HLC}}$ represents the high-level correlation (HLC)
contribution [those beyond CCSD(T)]. 
$ E_{\mathrm{HF}}^\infty$, $\Delta E_{\mathrm{CCSD(T)}}^\infty$,
and $E_{\mathrm{HLC}}$ can be 
obtained from HF and CC calculations
either using the small-core effective core potential (scECP) ECP28MDF
from Ref.~\cite{SmallECPStoll}, where the 4$s^2$4$p^6$5$s^1$ electrons of
\ce{Rb} are treated explicitly and all the others are modelled via a scalar-relativistic 
pseudopotential (PP), 
or using the all-electron spin-free exact two-component theory in its one-electron variant (SFX2C-1e) \cite{Dyall01,Liu09}
to treat scalar-relativistic effects. 
We can use spin-unrestricted (UHF) or spin-restricted open-shell (ROHF) approaches for the HF part of Eq.~\eqref{eq:HEATorig} and for generating the orbitals for the subsequent single-reference CC calculations. For the latter we used an unrestricted spin-orbital formalism in its singles and doubles variant augmented with a noniterative triples method based on a ROHF reference -- 
the ROHF-CCSD(T) method \cite{Raghavachari89,Bartlett90,Hampel92,Watts92} [also often referred to as `RHF-UCCSD(T)'].
The HLC contributions include the full triples correction obtained
as the difference between full CC singles doubles and triples (CCSDT) \cite{Noga87a,Scuseria88}
and CCSD(T) results using smaller basis sets and the quadruples correction obtained as
the difference between CC singles doubles triples augmented with noniterative quadruples [CCSDT(Q)]~\cite{Bomble05,Kallay05}.       
and CCSDT results (with, in general, even smaller basis sets). Here we used the CCSDT(Q)/B variant for ROHF reference~\cite{Kallay08}.

The overall goal of obtaining highly accurate ground-state PECs for \ce{Rb2+}
requires using large basis sets, which are flexible enough to describe both the
long-range part and the repulsive part as accurately as
possible. Moreover, the basis set should show smooth convergence behaviour when
extrapolating the results to estimate the CBS limit. Recently, 
correlation-consistent Gaussian basis sets
for use in correlated molecular calculations have been published for alkali
metal elements in Ref.~\cite{GrantHillBasis}. Those basis sets
[aug-cc-p(w)CV$n$Z-PP] are designed for the ECP28MDF
pseudopotential and are available up to quintuple-$\zeta$ quality. 
These methods define the theoretical framework for the present study. However, when proceeding to obtain the CCSD(T) part of the PECs for \ce{Rb2+},
a supposedly spurious hump in the long-range potential was observed. Therefore, no further effort was put into the calculation of the $\Delta E_{\mathrm{higher-rel}}$ part of Eq.~\eqref{eq:HEATorig}. Instead, our attention is placed on investigating
the origin of this problem in the CCSD(T) potential energy surface, for what
we have also carried out calculations of the \ce{Rb2+} PEC using approximate iterative triples methods
including CCSDT-n (n=1b, 2, 3, and 4) \cite{Lee84,Noga87b} as well as CCSDT and CCSDT(Q) \cite{Bomble05,Kallay05,Kallay08} calculations. 
Symmetry-broken ROHF-CCSD(T) calculations allowing for charge localization on one \ce{Rb} atom
have been carried out as well. 

The ECP-based ROHF-CCSD(T) calculations described above have been performed
using the \textsc{Molpro} 2018.2 program package 
\cite{MOLPRO-WIREs,MOLPRO_brief, OpenShellCCMolpro,OpenShellCCMolproErratum,UCCSDpTMolpro},
the SFX2C-1e-ROHF-CCSD(T), all CCSD(T)$_\Lambda$, \cite{Stanton97,Crawford98,Kucharski98} CCSDT, and CCSDT-n (n=1b, 2, 3, and 4) calculations have been carried out using the \textsc{cfour} program
package~\cite{cfour,Matthews20CFOUR, Watts92,Cheng11b}, and all CCSDT(Q) energies were computed using the
\textsc{mrcc} program suite~\cite{MRCCProg,Kallay01,Kallay05}.

\section{Theory}
\label{sec:theory}
Coupled-cluster theory is based on a similarity transformation of the Hamiltonian and a projective solution of the resulting stationary Schr\"odinger equation. This leads to the equations
\begin{subequations}
  \label{eq:CCeqs}
  \begin{align}
    E &= \Braket{\Phi_0|\bar{H}|\Phi_0} \label{eq:CCenergy}\\
    0 &= \Braket{\Phi_I|\bar{H}|\Phi_0}\,,
  \end{align}
\end{subequations}
where 
\begin{equation}
 \label{eq:SimTransHamil}
\bar{H} = e^{-\hat T}\hat H e^{\hat T}
\end{equation}
is the similarity transformed electronic clamped-nuclei Hamiltonian and 
 $\Ket{\Phi_0}$ the reference wavefunction. The excited determinants $\Ket{\Phi_I}$ are chosen to match the excitations
reached by he cluster operator  $\hat{T}=\sum_n\hat{T}_n$, which consists of $n$-fold excitation operators defined via
\begin{align}
\label{eq:ClusterOp}
   \hat{T}_n &= \frac{1}{(n!)^2}\sum_{ijk,\ldots}\sum_{abc,\ldots}[T_n]_{ijk\ldots}^{abc\ldots}\hat{a}_a^{\dagger}\hat{a}_b^{\dagger}\hat{a}_c^{\dagger}\cdots\hat{a}_k\hat{a}_j\hat{a}_i\cdots 
\end{align}
Here, the symobls $\hat{a}^{\dagger}$ are creation and the $\hat{a}$ annihilation operators, $[T_n]_{ijk\ldots}^{abc\ldots}$ are the cluster amplitudes  and the indices $a,b,c,\ldots$ run over virtual and $i,j,k,\ldots$ over occupied orbitals, respectively.
The electronic Hamiltonian $\hat{H}$ is usually formulated as
\begin{subequations}
  \label{eq:Hn}
  \begin{align}
    \hat{H} &= E_0 + \hat{f}_N + \hat{W}_N \\
    &= E_0 + \sum_{pq}f_p^q\hat{a}_p^\dagger\hat{a}_q + \frac{1}{4}\sum_{pqrs}g_{pr}^{qs}\hat{a}_p^\dagger\hat{a}_r^\dagger\hat{a}_s\hat{a}_q\,,
  \end{align}
\end{subequations}
with the one-particle operator $\hat{f}_N$ containing the Fock matrix $f_p^q$ and the
two-particle operator $\hat{W}_N$ with the anti-symmetrized two-electron
integrals $g_{pr}^{qs}$. For later reference, we note that the coupled-cluster equations, Eqs.~\eqref{eq:CCeqs}, can be summarized as an energy functional\cite{Arponen83}
\begin{equation}
\label{eq:Lambda}
    \mathcal L = \braket{\Phi_0|(1+\hat\Lambda)\bar{H}|\Phi_0}
\end{equation}
where the Lambda operator was introduced, consisting of a set of deexcitation operators with analogous definition to that of the excitation operators of the cluster operator.

For CCSD, the cluster operator is truncated after double excitations, but it is well-known that quantitative computations require at least an approximate account of triple excitations. In order to cut down the computational expense of full CCSDT computations, it is usual to approximate the triple excitations perturbatively.
One of the first approaches implemented is the CCSD[T] energy correction [originally called CCSD$+$T(CCSD)]~\cite{Urban1985}, which is based on a fourth-order perturbation theory contribution and is
given, assuming canonical (Hartree-Fock) orbitals, in terms of 
\begin{align}
  \label{eq:CCSDpT2}
  \Delta E^{[T]} &= \Delta E^{(4)} =
  -\braket{\Phi_0|\hat{T}_3^\dagger\hat{f}_N\hat{T}_3|\Phi_0} \notag \\ 
  &=
  -\frac{1}{36}\sum_{ijk}\sum_{abc}\left([T_3]_{ijk}^{abc}\right)^2\cdot
  D_{ijk}^{abc}\,,
\end{align}
with $\hat{T}_3$ defined via Eq.~\eqref{eq:ClusterOp} and $D_{ijk}^{abc}$ expressed in terms of orbital energies $\varepsilon_p$ via
\begin{align}
 \label{eq:CCSDpTdenominator}
   D_{ijk}^{abc} &= \varepsilon_a + \varepsilon_b + \varepsilon_c -
   \varepsilon_i - \varepsilon_j - \varepsilon_k 
\end{align}
The triples amplitudes are computed from the converged CCSD amplitudes
\begin{align}
  \label{eq:triplesAmp}
  \left[T_3\right]_{ijk}^{abc} &=\frac{-\braket{
      \Phi_{ijk}^{abc}|[\hat{W}_N,\hat{T}_{2,\text{CCSD}}]|\Phi_0}}{\varepsilon_a+\varepsilon_b+\varepsilon_c-\varepsilon_i-\varepsilon_j-\varepsilon_k}\,,
\end{align}
where $\Phi_{ijk}^{abc}$ denotes a triply excited determinant. By considering an additional term including CCSD singles excitations one obtains an energy correction, which is formally of fifth-order in the perturbation expansion, yielding
\begin{align}
  \label{eq:fifthorderenergy}
  \Delta E^{(5)} &= \Braket{\Phi_0|\hat{T}_1^\dagger\hat{W}_N\hat{T}_3|\Phi_0}\notag \\
  &= \frac{1}{4}\sum_{ijk}\sum_{abc}\,[T_1]_i^a\braket{bc||jk}[T_3]_{ijk}^{abc}\,.
\end{align}
The well-known CCSD(T) method~\cite{Raghavachari89,Watts92} includes this fifth-order term:
\begin{align}
  \label{eq:CCSDpT}
  \Delta E^{(T)} &= \Delta E^{(4)} + \Delta E^{(5)} = \Delta E^{[T]} + \Delta E^{(5)}\,.
\end{align}
The fifth-order term can be understood in terms of an alternative definition of the unperturbed system\cite{Stanton97} and was shown to be often essential for a good performance of the perturbative triples correction.\cite{Stanton1989}

In addition, a number of further approximations to CCSDT exist, which treat the triple excitations perturbatively, but include them self-consistently into the solution of the coupled-cluster equations. This is in particular the class of 
CCSDT-$n$ methods of which we in this work consider the variants $n$ = 1b, 2, 3, and 4.\cite{CCSDT-1-Impl,CCSDT-2-3-Impl,CCSDT-n_general,CCSDT-n_Urban} 
CCSDT-1b can be largely viewed as the self-consistent version of CCSD(T), as it includes the same leading-order terms in the coupled-cluster equations, which also lead to the perturbative energy expression, Eq.~\eqref{eq:CCSDpT}. In addition, it includes contributions of $\hat T_1\hat T_3$ to the doubles residual, which is thus complete (compared to the doubles residual of the full CCSDT method). The other methods, CCSDT-2 and CCSDT-3, subsequently include further terms in the residual for the triple excitations, while avoiding any $N^8$-scaling contributions. Hence, up to this point the only contribution of the cluster operator
$\hat{T}_3$ in the triple excitation residual appears via
$\braket{\Phi_{I_3}| \hat{f}_N\hat{T}_3|\Phi_0}$ defining an
equation  for determining the corresponding
$\hat{T}_3$ amplitudes independent of $\hat{T}_3$ itself. These amplitudes are
calculated `on the fly' immediately followed by calculating the
resulting contribution of $\hat{T}_3$ in the projection onto the singles
and doubles subspaces. These computational savings are lost when
proceeding to CCSDT-4 which partially includes $N^8$ terms. This method goes beyond the perturbative approximation of $\hat T_3$ and includes the full term $\braket{\Phi_{I_3}|[\hat{H},\hat{T}_3]|\Phi_0}$.
While the CCSDT-$n$ methods do not find wide use for the computation of ground state energies, they provide a useful hierarchy to investigate any short-comings of CCSD(T).

\begin{figure}[tb]
  \centering
  \includegraphics[width=\columnwidth]{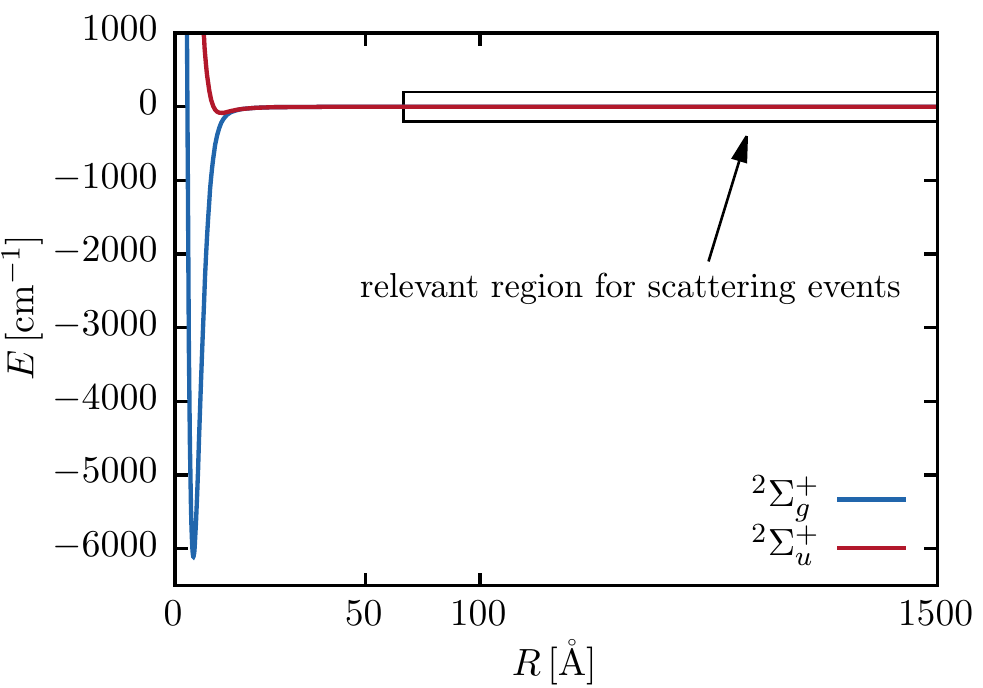}
  \caption{\label{fig:ExPEC}Schematic illustration of complete potential
    energy curves (PECs) of the \ce{Rb2+} ground states ${}^2\Sigma_g^+$ and
    ${}^2\Sigma_u^+$. For studying the ion-atom interaction we need high
    accuracy over the whole range of the PEC to obtain both the rovibrational
    structure and the long-range region, relevant for investigating scattering
    events, highly accurate.}
\end{figure}

\section{\label{sec:results}Results}

In this work we concentrate on the ROHF-CCSD(T) part of
Eq.~\eqref{eq:HEATorig} for the calculation of the
$\mathrm{X}{\,}^2\Sigma_g^+$ ground state of \ce{Rb2+}. There is a
second state of ungerade symmetry (i.e. $(1){\,}^2\Sigma_u^+$) which becomes 
degenerate to the former one in the long-range region, as shown in Fig.~\ref{fig:ExPEC}.
However, in the following we will solely focus on the long-range region of the
$\mathrm{X}{\,}^2\Sigma_g^+$ PEC.

\subsection{\label{sec:failure}Breakdown of CCSD(T)}
\begin{figure*}[tb]
  \centering
  \begin{minipage}[t]{.49\linewidth}
    \includegraphics[width=\columnwidth]{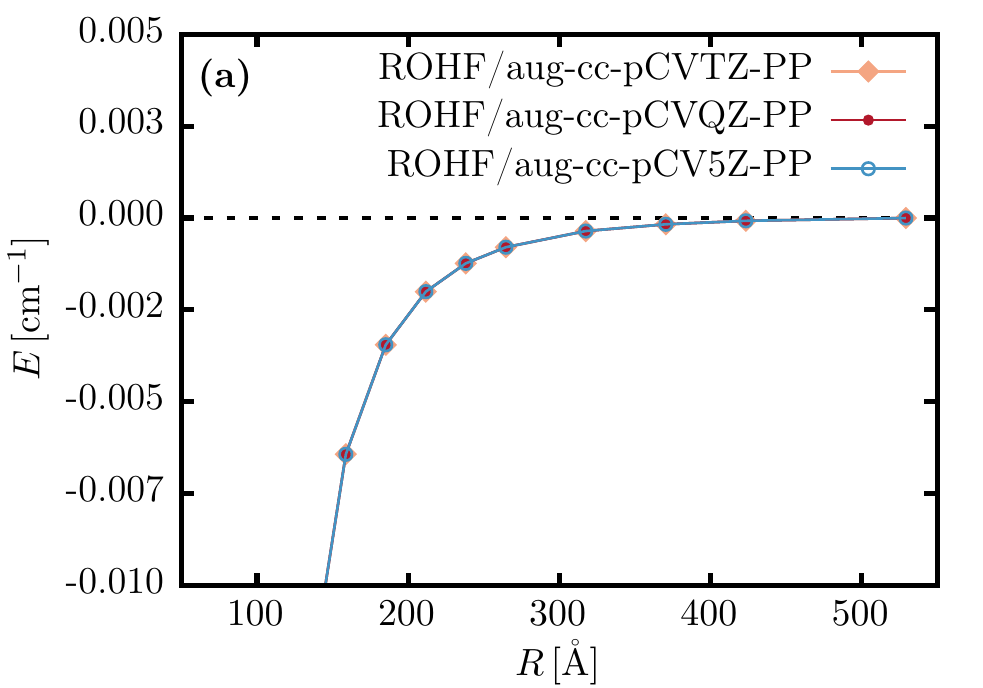}
  \end{minipage}
  \begin{minipage}[t]{.49\linewidth}
    \includegraphics[width=\columnwidth]{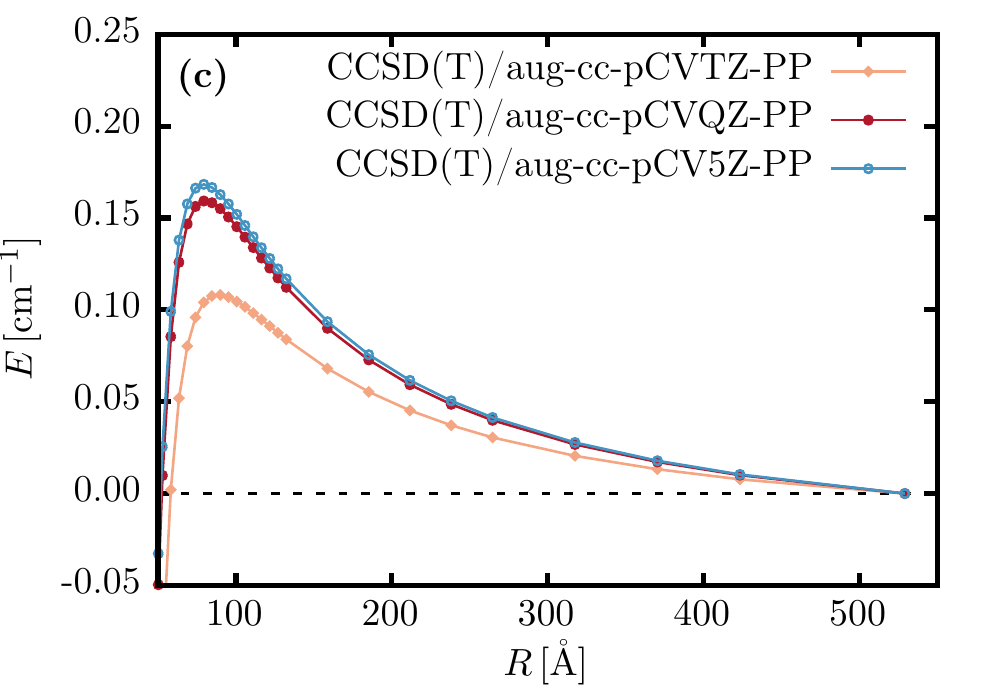}
  \end{minipage}
  \begin{minipage}[t]{.49\linewidth}
    \includegraphics[width=\columnwidth]{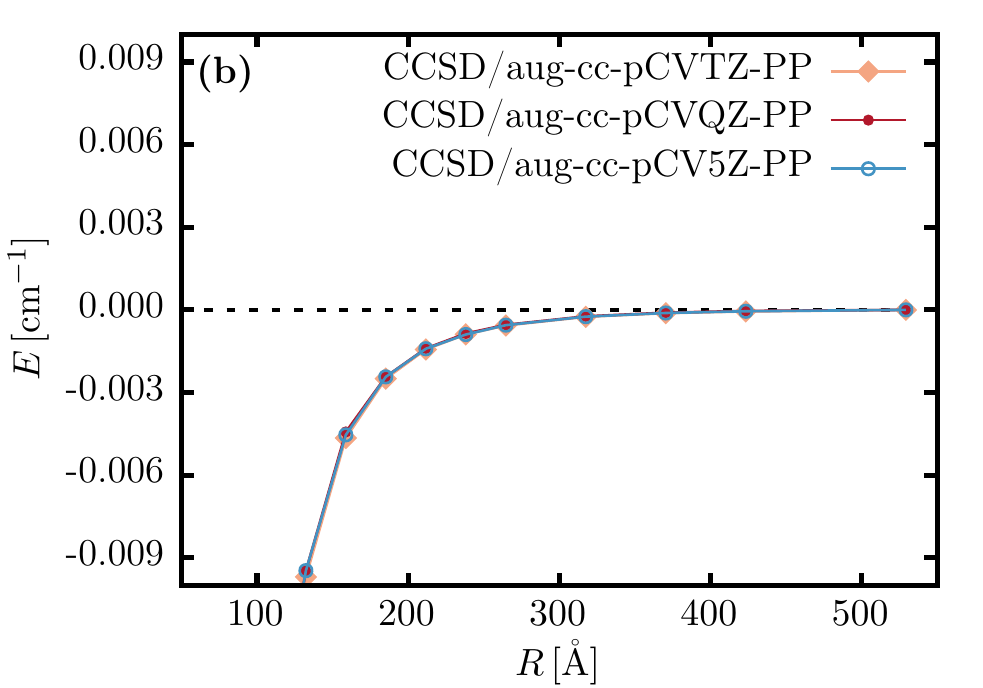}
  \end{minipage}
  \begin{minipage}[t]{.49\linewidth}
    \includegraphics[width=\columnwidth]{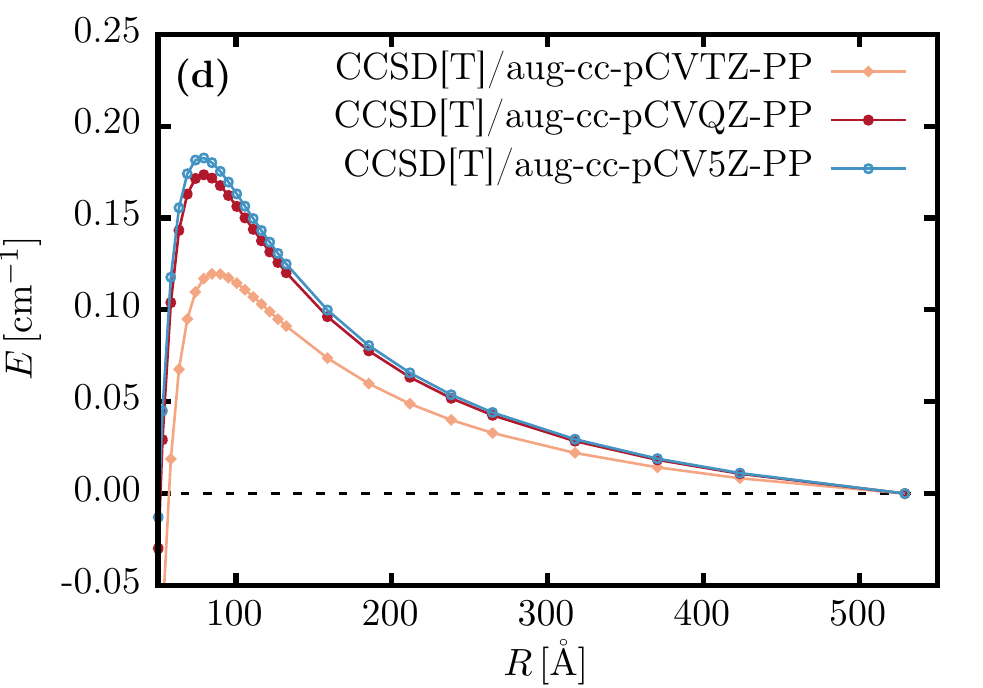}
  \end{minipage}
  \caption{\label{fig:basissetstogether}Overview of the long-range parts of
    the PECs of the \ce{Rb2+} ground state calculated at different levels of
    theory each using ECP28MDF. In~\textbf{(a)} the reference energies (ROHF)
    are shown for the aug-cc-pCV$n$Z-PP basis sets. In~\textbf{(b)} the
    coupled cluster energies with single and double excitations (CCSD) are
    shown. From Figs.~\textbf{(c)} and~\textbf{(d)} we obtain that including
    perturbative triples in the coupled cluster method either via (T) or [T]
    lead to unphysical humps in the long-range region. All energies were
    calculated with respect to the asymptote.}
\end{figure*}

In all calculations, we tightened the convergence thresholds for both the
underlying ROHF calculations and the subsequent
coupled-cluster part as much as possible to avoid numerical
errors. Figure~\ref{fig:basissetstogether} gives an overview of the resulting
long-range parts of the corresponding PECs for the aug-cc-pCV$n$Z-PP
basis sets ($n=$T, Q, 5). 
The curves 
for the ROHF reference and the CCSD energies show the expected long-range
behavior, i.e. a weakly attractive potential that decays in accordance with Eq.~\eqref{eq:LR}. However, including perturbative triples corrections via CCSD(T)
produces a small but clearly unexpected barrier in the long-range region at
$R\approx\SI{100}{\angstrom}$ with a magnitude of $\approx
0.15\,\mathrm{cm}^{-1}$ above the dissociation asymptote. The triples corrections included via [T] are correct to fourth order perturbation theory and produce a slightly more pronounced barrier at the same position. This indicates that the additional fifth-order terms included in the (T) correction serve for a slight, but certainly not sufficient, compensation.
This will be discussed in more detail in Sec.~\ref{sec:proof}. 

\begin{figure}[tb]
  \centering
  \includegraphics[width=\columnwidth]{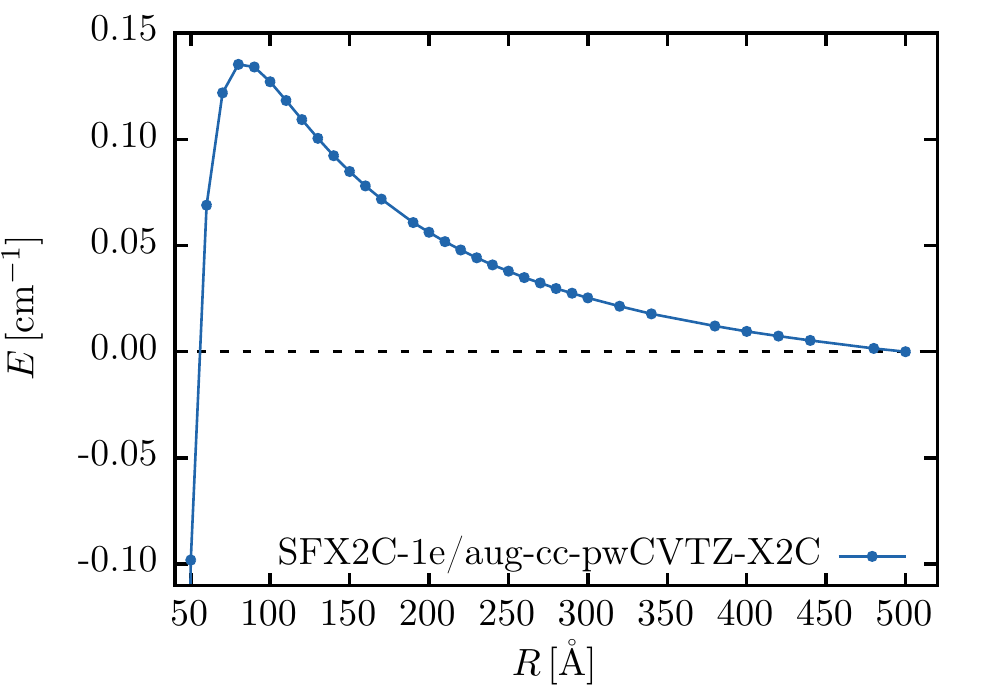}
  \caption{\label{fig:X2C}All electron SFX2C-1e calculation at
    ROHF-CCSD(T)/aug-cc-pwCVTZ-X2C level of theory. Here the scECP (ECP28MDF)
    is replaced by the X2C Hamiltonian thus correlating all electrons. Since
    we obtain the same long-range part as with ECP28MDF the hump is rather a
    general problem of the corresponding coupled cluster method than an
    artefact of the pseudopotential. The energies are calculated w.r.t. the
    last \emph{ab-initio} point.}
\end{figure}

To the best of our knowledge this kind of unphysical behavior seems to be
undocumented so far. It appears to be an inherent problem for the CCSD(T) method, since
other sources of error can be excluded after thoroughly investigating their
impact (see also supplementary material):
\begin{enumerate}[(i)]
\item numerical errors due to convergence issues: We used tightened thresholds a priori, with numerical noise for energies in the order of $<\mathcal{O}(2.5\cdot 10^{-7}\,\mathrm{cm}^{-1})$.
\item insufficient basis set: As seen in Fig.~\ref{fig:basissetstogether}, the shape of the PEC for ROHF-CCSD does not depend on the basis set and the height of the spurious barrier at the ROHF-CCSD(T) level even increases for larger basis sets. This also implies that basis set superposition is not a cause of the problem either, which we could also confirm by applying the counterpoise correction scheme to account for basis set superposition errors.
\item the choice of reference wavefunction: We computed the long-range tail of the PECs using UHF references with the \textsc{cfour} program and obtained virtually the same result, with absolute energy differences $\mathcal{O}(10^{-2}\,\mathrm{cm}^{-1})$.
\item use of spin-unrestricted or partially spin-restricted coupled cluster theory [i.e. RHF-UCCSD(T) or RHF-RCCSD(T)], see, e.g., Refs.~\citenum{OpenShellCCMolpro,OpenShellCCMolproErratum,UCCSDpTMolpro}: This only leads to energy differences in the long-range region in the order of $\mathcal{O}(10^{-4}\,\mathrm{cm}^{-1})$
\end{enumerate}
We also found that
this unphysical barrier is universal for \ce{X2+} systems (\ce{X} $=$ \ce{Li},\ce{Na},
\ce{K}, \ce{Rb}, \ce{Cs}), which is shown in the supplementary material. Moreover, it is not 
an artefact due to the approximative
nature of the scECP treatment 
since an all-electron SFX2C-1e-ROHF-CCSD(T) calculation at aug-cc-pwCVTZ-X2C level of theory 
leads to the long-range behavior
shown in Fig.~\ref{fig:X2C}.
Obviously, the long-range barrier is still present, at the same position with the same order
of magnitude. 
Finally, we note that there are no multireference effects expected for the \ce{Rb2+} system. The two near-degenerate states ($X{\,}^2\Sigma_g^+$ and $(1){\,}^2\Sigma_u^+$) are of
different symmetry and thus do not mix. This is in contrast to what has been reported, e.g., in Ref.~\citenum{Fedorov2014} for the ground state PEC of neutral \ce{LiNa}, where indeed multireference effects are present and CCSD(T) fails to correctly describe the bond cleavage.

\subsection{Higher excitations and iterative approximations}
CCSD[T] and CCSD(T) are non-iterative approximations to CCSDT. To further
investigate the origin of the long-range hump we also applied iterative
approximations to full CCSDT: the CCSDT-$n$, with $n=$ 1b, 2, 3, 4,
methods~\cite{CCSDT-n_general,Lee84,Noga87b}. We used
the ECP28MDF pseudopotential and the aug-cc-pCVTZ-PP basis
set in these calculations. 

As outlined in Sec.~\ref{sec:theory} these methods include contributions due to triples excitations conveyed via $\hat{T}_3$ into the solution of the coupled cluster equations. Here all approaches, except the CCSDT-4 method, avoid including any terms with $N^8$ scaling.

\begin{figure*}[tb]
  \centering
  \begin{minipage}[t]{.49\textwidth}
    \includegraphics[width=\columnwidth]{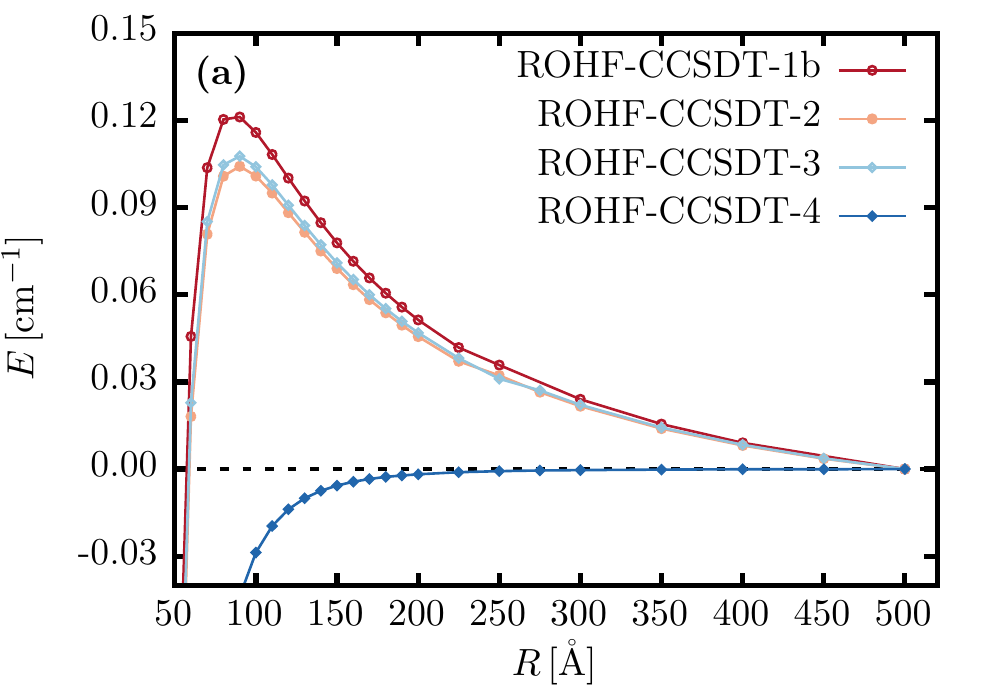}
  \end{minipage}
  \begin{minipage}[t]{.49\textwidth}
    \includegraphics[width=\columnwidth]{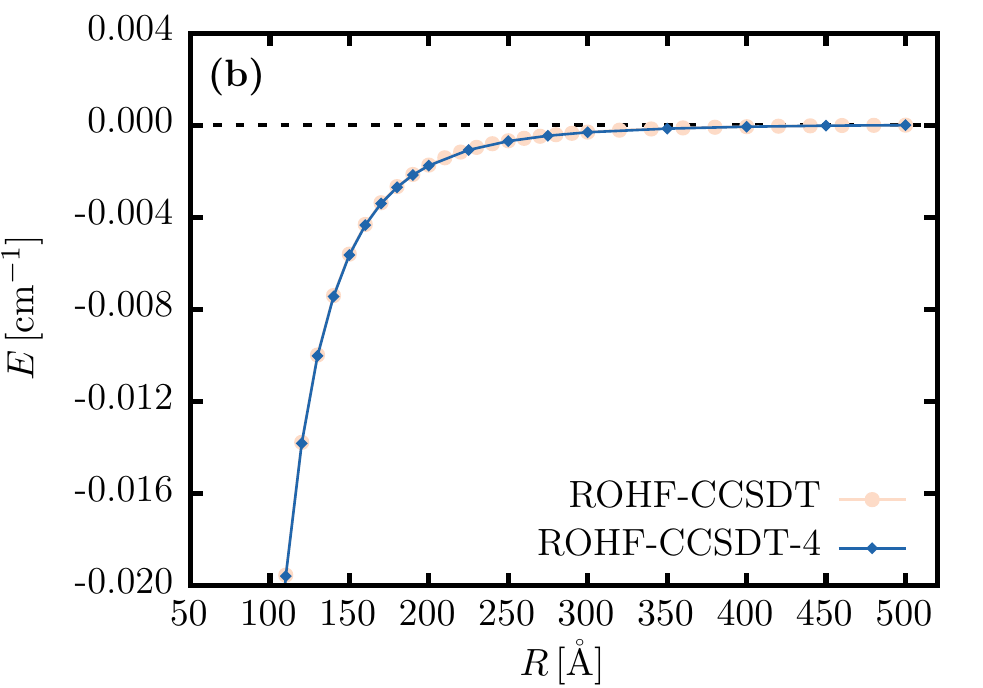}
  \end{minipage}
  \caption{\label{fig:CCSDTandCCSDT-n}\textbf{(a)} Long-range
    part of the interaction energies 
    using different iterative approximations to CCSDT.  \textbf{(b)} Comparison of CCSDT-4 and full
    CCSDT interaction energies. All computatations use a ECP28MDF pseudopotential, a 
    aug-cc-pCVTZ-PP basis set, and an ROHF reference. }
\end{figure*}

The resulting long-range PECs of these iterative approximations to CCSDT
are shown in Fig.~\ref{fig:CCSDTandCCSDT-n}~(a).
Again, we obtain a hump for CCSDT-1b, CCSDT-2, CCSDT-3 at the same position
($\approx\SI{100}{\angstrom}$) and of the same magnitude ($\approx
0.1\,\mathrm{cm}^{-1}$) as we have already seen for the non-iterative
methods. Including more terms in the approximation scheme leads to a decrease
in the size of the bump. However, only with the inclusion of the full
Hamiltonian in the projection onto the excited triples manifold, i.e. for
CCSDT-4, the artificial barrier disappears. But, as already mentioned this method is already as
expensive as the full CCSDT calculation. The resulting long-range tails are presented in
Fig.~\ref{fig:CCSDTandCCSDT-n}~(b) with both methods leading to the same shape in the asymptotic region.

This suggests the hypothesis that the
$\braket{\Phi_0|\hat{T}_3^\dagger\hat{f}_N\hat{T}_3|\Phi_0}$ term, shared by
all problematic methods (CCSD[T]/CCSD(T) as well), does not correctly account
for interatomic interactions. 
In fact, this term only contains the interaction with the Hartree-Fock density of the other atom.

\begin{figure}[tb]
  \centering
  \includegraphics[width=\columnwidth]{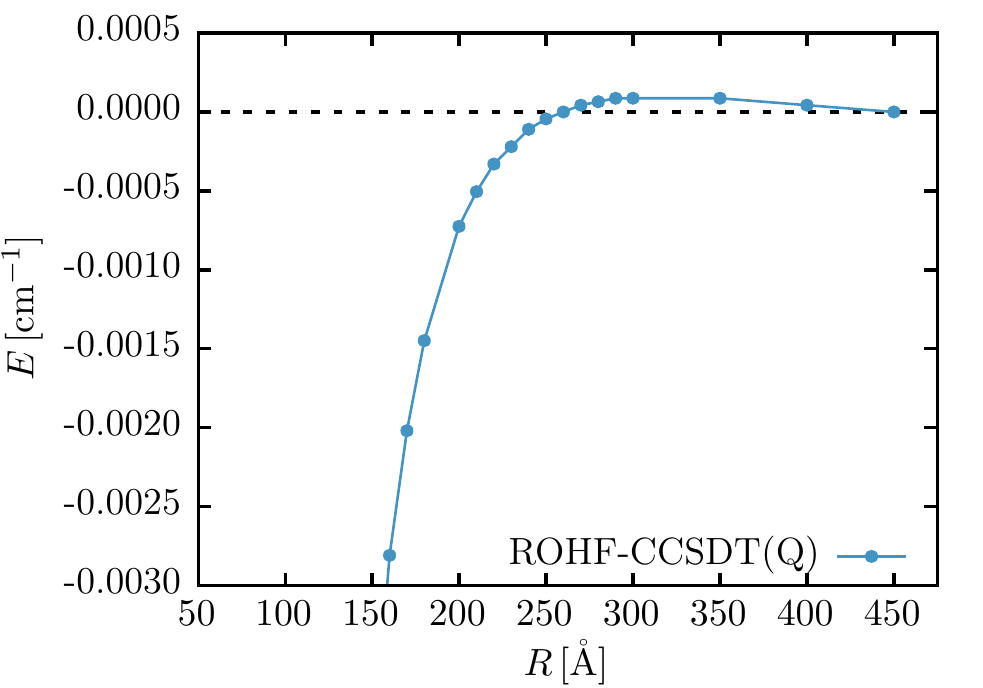}
  \caption{\label{fig:CCSDTpQ}ROHF-CCSDT(Q) long-range part of the interaction
    energy w.r.t. the last \emph{ab-initio} point calculated with the
    \textsc{mrcc} program suite using ECP28MDF/aug-cc-pwCVTZ-PP. The hump
    moved to the right compared to UCCSD(T) and the iterative approximations
    to CCSDT. It is smaller in size since (Q) corrections are in general
    smaller than (T) ones.}
\end{figure}

We obtain the same phenomenon if we perturbatively include even higher
excitations such as CCSDT(Q). The corresponding result is shown in
Fig.~\ref{fig:CCSDTpQ}.
The hump is smaller in size, as the contributions of connected quadruples are generally smaller than those of connected triples. Nevertheless, despite its smallness the artificial barrier completely spoils the long-range behavior of the potential, which is important for correct predictions on the scattering physics. 

\subsection{\label{sec:symmetrybreaking}Symmetry breaking}
In general \ce{X2+} -- systems, with \ce{X} $=$
\ce{Li}, \ce{Na}, \ce{K}, \ce{Rb}, \ce{Cs}, are characterized by the point group
$D_{\infty h}$.
This implies the asymptotical
indistinguishability of \ce{Rb+}$+$\ce{Rb} and \ce{Rb}$+$\ce{Rb+}, which is also clear from a fundamental quantum mechanical point of view. The correct asymptotic behavior is given in terms of a superposition of both limiting cases, i.e.
\begin{subequations}
  \label{eq:SuperposStates}
  \begin{align}
    \ket{X{\,}^2\Sigma_g^+} &= \frac{1}{\sqrt{2}}\left(\ket{0,+} +
      \ket{+,0}\right) \\
      \ket{(1){\,}^2\Sigma_u^+} &= \frac{1}{\sqrt{2}}\left(\ket{0,+} - \ket{+,0}\right)\,.
  \end{align}
\end{subequations}
The zeroth-order description of the system is a mean-field approximation (Hartee-Fock), which involves the self-consistent-field (SCF) solution 
for the corresponding equations. This need for self-consistent solutions leads to different orbitals for \ce{Rb+} and \ce{Rb} and with that the solution of the separated fragments is at conflict with the symmetry requirement that the two cases \ce{Rb+}$+$\ce{Rb} and vice versa are quantum-mechanical indistinguishable. All the orbitals of \ce{Rb} and \ce{Rb+} are a ``compromise'' of the neutral and ionic orbitals.
The mean-field solution also defines the Fockian, the effective one-electron potential of the system. It plays an important role for defining perturbative approximations in the coupled-cluster equations. Here, rather than describing the correct superposition, it contains the compromise solution with half an electron on the right and half an electron on the left side, possibly explaining the repulsive long-range barrier.

\begin{figure*}[tb]
  \centering
  \begin{minipage}[t]{.49\linewidth}
    \includegraphics[width=\columnwidth]{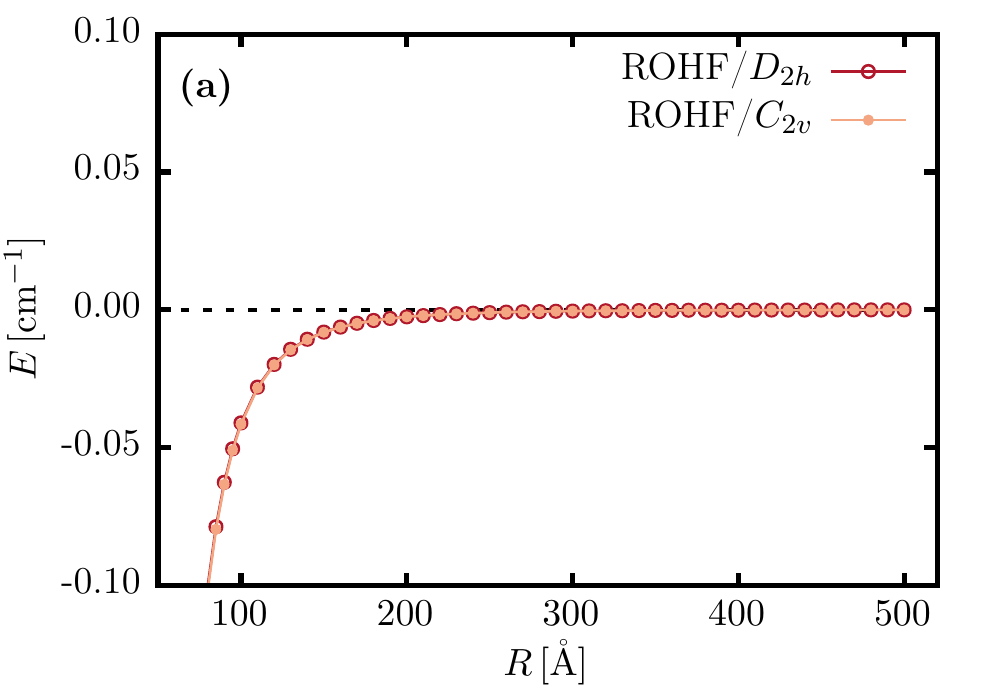}
  \end{minipage}
  \begin{minipage}[t]{.49\linewidth}
    \includegraphics[width=\columnwidth]{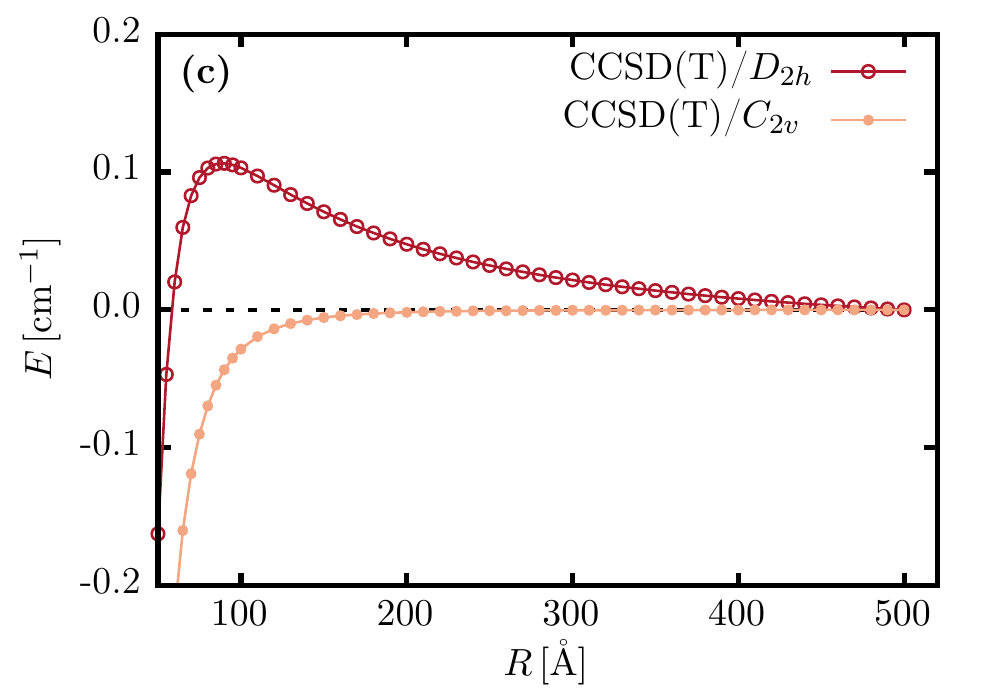}
  \end{minipage}
  \begin{minipage}[t]{.49\linewidth}
    \includegraphics[width=\columnwidth]{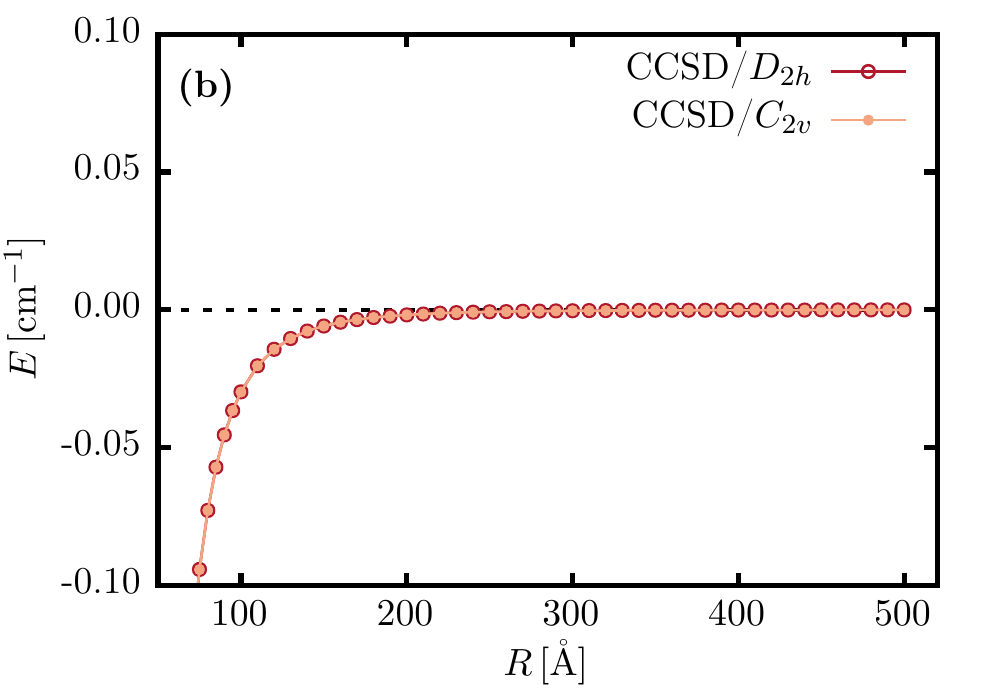}
  \end{minipage}
  \begin{minipage}[t]{.49\linewidth}
    \includegraphics[width=\columnwidth]{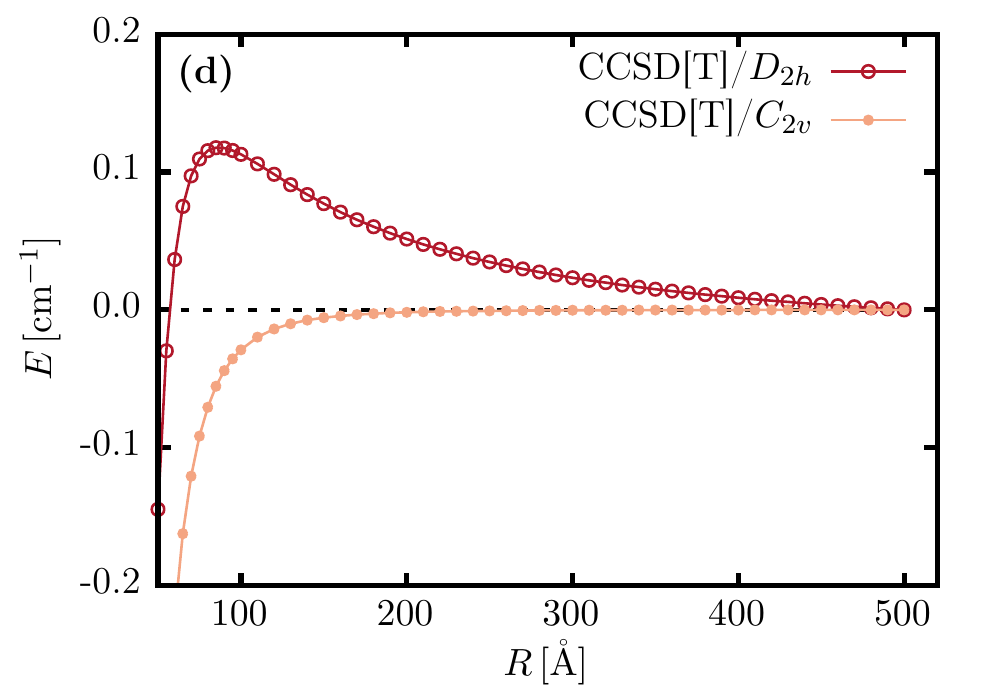}
  \end{minipage}
  \caption{\label{fig:OverviewC2vD2h}Comparison of the long-range parts using
    either symmetry-adapted ($D_{2h}$) or symmetry-broken ($C_{2v}$) ROHF orbitals. In~\textbf{(a)} the reference (ROHF) interaction energies are
    given w.r.t. the asymptote. In~\textbf{(b)} the CCSD
    results are depicted, while~\textbf{(c)} and~\textbf{(d)} corresponding to
    CCSD(T), respectively CCSD[T] interaction energies. All results were
    computed using ECP28MDF/aug-cc-pCVTZ-PP.}
\end{figure*}

If this is true, breaking the system's symmetry should lead to a barrierless asymptotic
behavior and thus size consistent mean-field solutions should be elements of the point group $C_{\infty v}$.
Quantum mechanically speaking we project on one of the two
limiting cases ($\ket{0,+}=\ce{Rb}+\ce{Rb+}$ or vice versa $\ket{+,0}$). 
To test this hyothesis, we carried out CCSD, CCSD[T] and CCSD(T) computations using symmetry-broken ROHF orbitals. The resulting long-range tails of the PECs
are given in Fig.~\ref{fig:OverviewC2vD2h}.
The results from Sec.~\ref{sec:failure} shown in
Fig.~\ref{fig:basissetstogether} are given for comparison.

The results demonstrate that the long-range hump can be avoided by reducing the symmetry to $C_{\infty v}$. At short-range these symmetry-broken solutions 
collapse to the symmetric one. This is illustrated in more detail in the
supplementary material. 
The findings suggest that the best model for the long-range region is based on the symmetry-broken solutions. 
So far the most promising approach is to use symmetry-broken (T) and (Q) corrections for the long-range tail and properly merge with symmetry-adapted solutions for smaller internuclear distances. With this all terms in the additivity scheme according to 
Eq.~\eqref{eq:HEATorig} 
are well-defined paving the way for a highly accurate PEC. The details on this will be published in a subsequent study.

\section{\label{sec:proof}Discussion}
Our findings so far suggest that the physical origin of the artificial
long-range humps is connected with the underlying mean-field character of our
calculations and the way it enters coupled cluster approximations.
To fully understand the problem and to
reveal the source of error in approximative coupled cluster methods we have to
go further into the theory and investigate the coupled cluster
equations~\eqref{eq:CCeqs} in more detail.

\subsection{Qualitative Analysis}

\begin{figure*}[tb]
    \centering
    \begin{minipage}[t]{.49\linewidth}
    \includegraphics[width=\columnwidth]{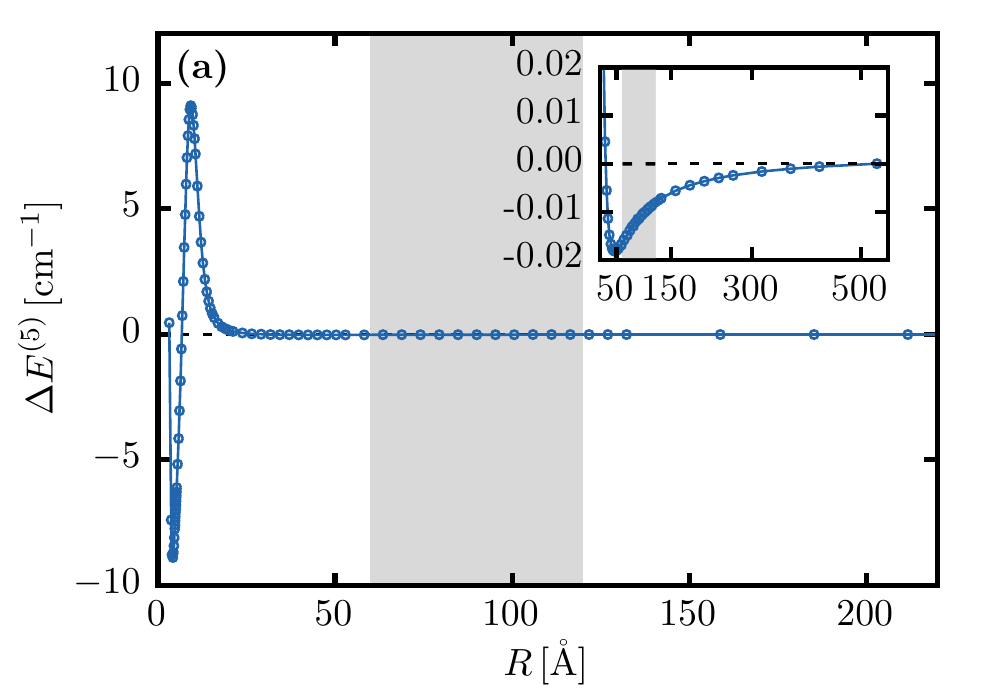}
    \end{minipage}
    \begin{minipage}[t]{.49\linewidth}
    \includegraphics[width=\columnwidth]{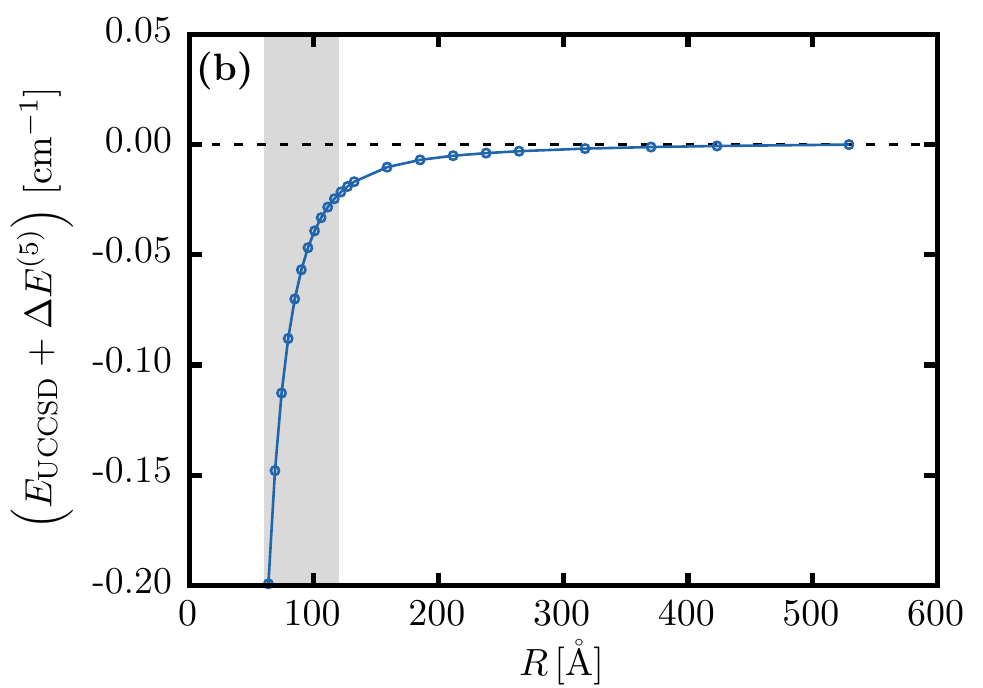}
    \end{minipage}
    \caption{\label{fig:AnalyzeCCeqs}Visulization of the impact of the fifth-order perturbative triples correction as non-iterative approximation to CCSDT. The pure fifth-order energy term according to
    Eqs.~\eqref{eq:fifthorderenergy} is given
    in~\textbf{(a)} relative to the asymptote. Adding
    $\Delta E^{(5)}$ to the CCSD energy leads to the correct long-range
    behaviour as shown in~\textbf{(b)}. The areas highlighted in gray mark the position of the long-range hump, where the inset in~\textbf{(a)}
    shows a zoom into this area.}
\end{figure*}

The $E^{(5)}$ energy contribution is shown in
Fig.~\ref{fig:AnalyzeCCeqs}~(a) with the area usually
containing the artificial long-range barrier highlighted in gray. The inset
demonstrate that in this region 
the fifth-order term is purely attractive.
This finally explains our observations from Fig.~\ref{fig:basissetstogether} of Sec.~\ref{sec:failure} showing a slightly more pronounced long-range hump for [T] corrections compared to (T) ones. 
Furthermore, as shown in Fig.~\ref{fig:AnalyzeCCeqs}~(b), we can conclude that the pure fifth-order correction is not connected to the discussed problem.
However, a priori it is still not clear
whether the nature of this phenomenon is due the approximation in the energy
expression or due to the use of approximate triples amplitudes.
The latter can be tested rather systematically by using
$\hat{T}_3$ amplitudes from CCSDT in Eq.~\eqref{eq:CCSDpT2}. This will be
investigated in the following.

\subsection{Straightforward improvement of the triples correction}

\begin{figure*}[tb]
  \centering
  \includegraphics[width=.8\linewidth]{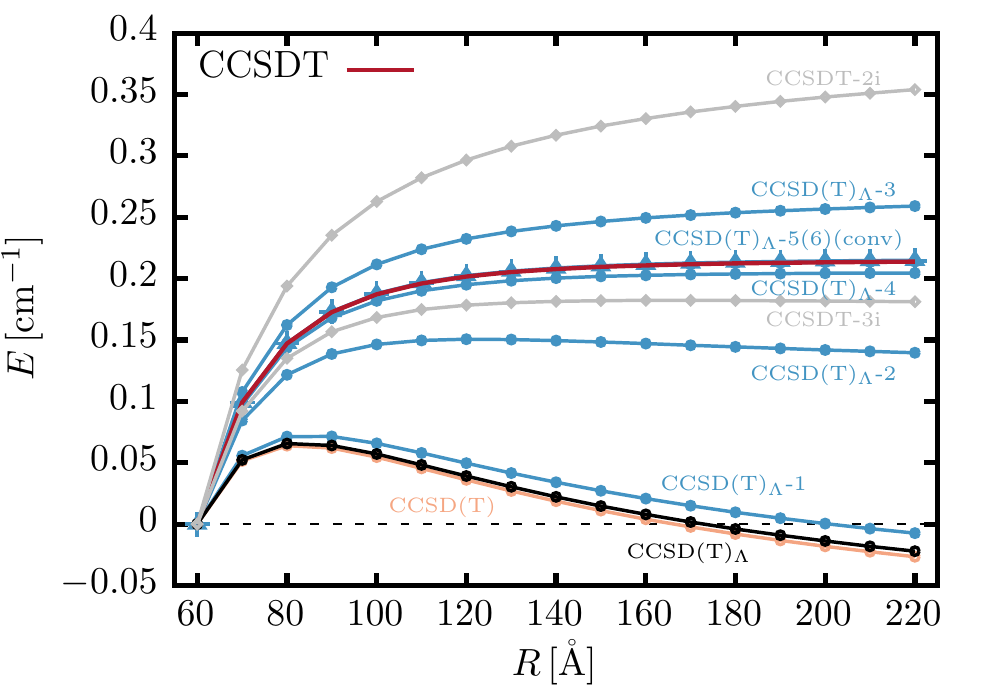}
  \caption{\label{fig:CCSDpT_Lambda-n}Comparison of potential energy curves of \ce{Rb2+}
    calculated with ECP28MDF/aug-cc-pwCVTZ-PP and different levels of the
    CCSD(T)$_\Lambda$-n method introduced in this work (see text for more
    details). The method refers to the use of $\hat{T}_3$ amplitudes obtained
    at the m'th CCSDT iteration. CCSD(T)$_\Lambda$-conv denotes the use of
    converged triples amplitudes. CCSDT-2i(3i) is the CCSDT energy after the
    second (third) iteration. All PECs are plotted w.r.t. $R=\SI{60}{\angstrom}$.}
\end{figure*}

To understand the appearance of artificial humps in the CCSD(T)
curves, we carried out calculations using a closely
related formulation for the triples contribution based on analytic energy
derivative formulation of CCSD. \cite{Adamowicz84,Scheiner87}  A CCSD Lagrangian \cite{Arponen83,Helgaker88} with triple excitations
treated as an ``external'' perturbation with a fictitious field strength
$\chi$ is used here
\begin{align}
  \label{Eq1}
  \mathcal{L}(\hat{T}_1,\hat{T}_2, \chi \hat{T}_3) &=
  \braket{0|(1+\hat{\Lambda}_{\text{CCSD}})\bar{H}[\chi]|0}\,,
\end{align}
in which $\hat{\Lambda}_{\text{CCSD}}=\hat{\Lambda}_{1, \text{CCSD}}+\hat{\Lambda}_{2, \text{CCSD}}$ 
represents the CCSD $\Lambda$ operator, and 
$\hat{T}_3$ contributes to $\bar{H}$, defined in Eq.~\eqref{eq:SimTransHamil}, 
in the same way as in the CCSDT method. 
Given the exact $\hat{T}_3$, 
a finite-field CCSD calculation defined as the left-hand side of Eq. (\ref{Eq1}) 
with $\chi=1$ produces the exact CCSDT energy. This finite field CCSD calculation uses the CCSDT equation, but treats the triples as the perturbation.
Hence the CCSDT energy can be expanded in a Taylor series 
\begin{align}
  \label{second}
  E_{\text{CCSDT}} &=\mathcal{L}(\hat{T}_1,\hat{T}_2,\chi \hat{T}_3)\Big|_{\chi=1}
  \notag \\
  &= \Big[\mathcal{L}\Big|_{\chi=0}
    +\chi \frac{\diff {\mathcal{L}}}{\diff \chi}\Big|_{\chi=0}
    \notag\\&+\frac{1}{2} \chi^2 \frac{\diff^2 {\mathcal{L}}}{\diff \chi^2}\Big|_{\chi=0}
    +\cdots\Big]_{\Big|_{\chi=1}} \notag \\
  &= \mathcal{L}\Big|_{\chi=0}
  + \frac{\diff {\mathcal{L}}}{\diff \chi}\Big|_{\chi=0}
  +\frac{1}{2} \frac{\diff^2 {\mathcal{L}}}{\diff \chi^2}\Big|_{\chi=0}
  +\cdots, 
\end{align}
with the unperturbed energy being the CCSD energy 
\begin{align}
	\mathcal{L}\Big|_{\chi=0} &= E_{\text{CCSD}} 
\end{align}
and the first-order correction given by
\begin{align}
  \label{first} 
  \frac{\diff {\mathcal{L}}}{\diff \chi}\Big|_{\chi=0} &= 
  \Braket{0|\hat{\Lambda}_{\text{CCSD}}\frac{\partial \bar{H}[\chi]}{\partial \chi}|0}_{\Big|_{\chi=0}} \notag \\
  &= \braket{0|\hat{\Lambda}_{\text{CCSD}} [\hat{H}, \hat{T}_3]|0}\,. 
\end{align}
Eq. (\ref{first}) offers a flexible framework for obtaining leading triples
corrections to the CCSD energy.  When the leading-order contribution to $\hat{T}_3$
is used, i.e. the triples amplitudes defined in Eq.~\eqref{eq:triplesAmp},
Eq. (\ref{first}) reduces to the CCSD(T)$_\Lambda$ method originally derived
within the equation-of-motion CC framework. \cite{Stanton97,Crawford98,Kucharski98} 
Eq. (\ref{first}) is also
compatible with the use of improved $\hat{T}_3$ amplitudes obtained from
iterative solutions of the CCSDT amplitude equations. This is particularly
useful for the present purpose to understand whether the small humps in the
CCSD(T) 
potential energy curves originate from the
approximation in the energy expression or from the approximation of using
leading-order $\hat{T}_3$.  A straightforward approach to improve $\hat{T}_3$
is to solve the CCSDT amplitude equations iteratively. Here we define
$E_{\text{T}}{\text{[CCSD(T)}}_\Lambda\text{-n}]$ as the triples energy
correction using Eq. (\ref{first}) with $\hat{T}_{3,\text{n'th}}$ obtained
from the n'th iteration of CCSDT equations with converged CCSD amplitudes
adopted as the initial guess
\begin{align}
  E_{\text{T}}{\text{[CCSD(T)}}_\Lambda\text{-n}] &=
  \braket{0|\hat{\Lambda}_{\text{CCSD}} [\hat{H}, \hat{T}_{3,\text{n'th}}]|0}.
  \label{DN}
\end{align}
The use of $T_3$ from a converged CCSDT calculation provides the first-order correction  
from the triples amplitudes to the CCSD energy
\begin{align}
  E_{\text{T}}{\text{[CCSD(T)}}_\Lambda\text{-conv}]=
  \braket{0|\hat{\Lambda}_{\text{CCSD}} [\hat{H}, \hat{T}_{3,\text{CCSDT}}]|0}.
  \label{full}
\end{align}
Here ``$\text{-conv}$'' denotes the use of converged CCSDT $\hat{T}_3$. 
Although it is obviously not practically useful, 
Eq. (\ref{full}) defines the limit of the accuracy that can be obtained using Eq. (\ref{first}).

The CCSD(T)$_\Lambda$-n methods are related to available methods for obtaining triples corrections, \cite{Hirata01,Wloch05,Wloch06a,Wloch06b,Eriksen14a,Eriksen14b,Eriksen15} the most intimately to the CCSD(T-n) methods \cite{Eriksen14a,Eriksen14b,Eriksen15} derived using CCSD Lagrangian, 
which also treats CCSD as the unperturbed state. 
The difference lies in that the 
schemes outlined so far refrain from performing an M{\o}ller-Plesset perturbation analysis
and limit the consideration to the first derivative with respect to the triples amplitudes,
whereas, the CCSD(T-n) derivation expands 
the Lagrangian order by order in terms of the fluctuation potential. 
Since the first iteration of CCSDT amplitude equations in CCSD(T)$_\Lambda$-n
uses CCSD solutions as the initial guess and 
generates first-order $\hat{T}_3$. Substituting this $\hat{T}_3$ into Eq. (\ref{first}) 
gives a triples correction correct to second-order, i.e., CCSD(T)$_\Lambda$-1
appears to be equivalent to CCSD(T-2).
CCSD(T-4) contains contributions from higher order, i.e., the second term in Eq. (\ref{second}), which 
is not considered in CCSD(T)$_\Lambda$-n. 
The CCSD(T-n) methods are more efficient, since no storage of T$_3$ is needed.
However, the CCSD(T)$_\Lambda$-n approaches may have the advantage that
the use of convergence-acceleration techniques during iterative solutions
of CCSDT equation such as direct inverse of iterative space (DIIS) \cite{Pulay80}
may smooth the convergence when the plain iterative solutions
exhibit an oscillating behavior.

We have performed $\text{CCSD(T)}_\Lambda\text{-n}, n=1-6$ and
$\text{CCSD(T)}_\Lambda\text{-}\infty$ calculations for potential energy
surfaces of Rb$_2^+$ using aug-cc-pwCVTZ basis set.  As shown in
Fig.~\ref{fig:CCSDpT_Lambda-n}, the $\text{CCSD(T)}_\Lambda\text{-n}$ results
systematically converge to the $\text{CCSD(T)}_\Lambda\text{-}\infty$ results,
which is essentially indistinguishable from CCSDT results.  The
$\text{CCSD(T)}_\Lambda\text{-1}$ curve shows an artificial hump similar to
the case of $\text{CCSD(T)}_\Lambda$.  The hump is significantly reduced in
the $\text{CCSD(T)}_\Lambda\text{-2}$ curve and is eliminated using methods
with more than two iterations. These results clearly support that the
artificial hump in the CCSD(T)
curve originates from the
approximation of $\hat{T}_3$ This is
also true for iterative approximations to CCSDT, i.e. CCSDT-$n$ ($n=1b,2,3$)
since the equation for determining triples amplitudes approximately is
formally similar to Eq.~\eqref{eq:triplesAmp}. 
The physical explanation
behind this is based on the underlying mean-field approximation and the need to define self-consistent solutions leading to different orbitals for \ce{Rb+} and \ce{Rb} which has been already discussed in detail in Sec.~\ref{sec:symmetrybreaking}.

Note that the $\text{CCSD(T)}_\Lambda\text{-n}$ energies converge more rapidly
than CCSDT energies with respect to iterative solution of CCSDT amplitude
equations, as also demonstrated in Fig.~\ref{fig:CCSDpT_Lambda-n}. The
$\text{CCSD(T)}_\Lambda\text{-n}$ triples correction is less sensitive to the
quality of $\hat{T}_3$, since it is only a small fraction of the total CCSDT
energy.  Finally, we note that, while the $\text{CCSD(T)}_\Lambda\text{-n}$
methods have proven useful in the present context, the potential usefulness in
calculations of chemical properties remains to be explored.

\section{\label{sec:ConclusionOutlook}Conclusion}
This work shows that several standard coupled-cluster methods with noniterative or approximative iterative treatments of triple excitations can lead to unphysical potential energy curves for \ce{X2+} systems, $\ce{X}\in\lbrace \ce{Li},\ce{Na},\ce{K},\ce{Rb},\ce{Cs}\rbrace$, with a repulsive long range barrier. Although this effect is in the order $\mathcal{O}(10^{-1}\,\mathrm{cm}^{-1})$ it would lead to severe problems when using the corresponding PECs for highly accurate studies in the context of ultracold chemistry. We unraveled the origin of this phenomenon by studying the ground state PEC of \ce{Rb2+}. It arises from the need to define self-consistent solutions which at the same time cannot be both consistent with the separated fragments (different orbitals for \ce{Rb+} and \ce{Rb}) and with the quantum mechanically imposed symmetry requirement (indistinguishable cases \ce{Rb+}$+$\ce{Rb} and \ce{Rb}$+$\ce{Rb+}). This problem lives on in the Fockian and affects the perturbative estimates of the $\hat{T}_3$ amplitudes, which finally lead to the wrong behavior of the PEC.
This was demonstrated quantitatively by using a new ``CCSD(T)$_\Lambda$-n'' scheme.

For the \ce{Rb2+} molecule we found that symmetry-broken CCSD(T) solutions lead to physically correct long-range PECs while symmetry-broken and non-broken CCSD curves virtually coincide in this region. From this we conclude that (T) corrections from symmetry- broken calculations can be used for estimating the complete basis set limit of the long-range part of the PEC. In the same way we could proceed with (Q) corrections and smaller basis sets eventually defining a protocol for obtaining a highly accurate global PEC for the ground state of \ce{Rb2+}. This will be thoroughly investigated in a subsequent study.

\section*{Supplementary Material}

See supplementary material for technical details on the independence of the
long-range hump on possible sources of error, for the universality of the
current problem for \ce{X2+} systems and for more details on symmetry
breaking.

\begin{acknowledgments}
  J.S. and A.K. would like to acknowledge funding by
  IQ\textsuperscript{ST}. The research of IQ\textsuperscript{ST} is
  financially supported by the Ministry of Science, Research, and Arts
  Baden-Württemberg. The work at Johns Hopkins university has been supported
  by the National Science Foundation, under grant No. PHY-2011794.
\end{acknowledgments}

\section*{Data Availability}

The data that support the findings of this study are available within this
article and its supplementary material.

\providecommand{\noopsort}[1]{}\providecommand{\singleletter}[1]{#1}%

\end{document}